# Advancing optical imaging systems with digital fabrication

Tobias Wenzel[1,2]*, Richard Bowman[3], Gemma S. Cairns[4], Benedict Diederich[5], Matias Hurtado[1], Vicente Parot[1], Vittorio Saggiomo[6]

## Abstract

Optical imaging technologies are central to discovery in the life and physical sciences, yet their impact depends on how readily they can be built, adapted, and sustained across laboratories. Digital fabrication, including desktop 3D printing, offers new ways to engineer imaging instruments by simplifying assembly, lowering replication barriers, and enabling modular integration and local refinement. Here we examine, using open microscopy as a transparent case, how digitally fabricated components support adaptable, research-grade optical systems while enabling faster innovation cycles and distributed improvement. We outline practical design guidelines and discuss emerging developments that may further advance accessible, high-performance imaging.

## Introduction

As new optical imaging methods are continuously developed, the ability to deploy, adapt, and sustain experimental instrumentation is becoming a key factor in how quickly techniques spread across laboratories and research areas. While material cost is important when sharing instrumentation designs, dissemination is often limited by complexity, e.g. multi-component assemblies, alignment sensitivity, supplier dependencies, and the challenges of modification and maintenance after systems are built. Consequently, the broader impact of new imaging technologies depends not only on their optical performance, but also on how easily systems can be acquired, replicated, modified, and recombined to evolve across laboratories. However, there is still a lack of well-defined engineering strategies that enable rapid and accessible innovation of optical imaging systems with reliable replication and reproducibility.

Digital fabrication, especially desktop 3D printing, provides an opportunity to reduce both cost and complexity of optomechanical assemblies by enabling rapid online design sharing with the agility of digital data, distributed local fabrication, modular integration, and user-driven modification[1–4]. When applied strategically, digitally fabricated components and maker tools can function as final parts of research-grade instruments to support compact integration and simplify alignment, maintenance, and upgrades. These capabilities enable faster and more user-centered design iteration cycles, facilitate replication, reproducibility and feasibility across different locations, and reduce dependence on narrow supplier ecosystems, particularly in settings with limited access[5] to specialized commercial components.

Digital fabrication in scientific instrumentation is a relatively recent but rapidly expanding approach[5,6] for innovating[7] and disseminating research technology. The most widely adopted method in imaging research is plastic filament 3D printing[8], commonly referred to as fused deposition modelling (FDM) or fused filament fabrication (FFF). Access to consumer-grade desktop 3D printers and distributed fabrication workflows has expanded substantially in recent years[2,9], reflecting the widespread adoption of low-cost additive manufacturing across regions[6]. Accessible workflows also include stereolithographic resin printing (SLA, mSLA, and DLP), laser cutting of planar components, and desktop CNC milling of polymers or metals. More advanced digitally controlled methods, such as PCB manufacturing, powder-based printing, or precision machining services, are increasingly integrated through outsourcing. Combined with affordable microcontrollers, single-board computers, industrial motors, and machine vision components, these fabrication tools have substantially expanded the capacity of small teams to design, replicate, and refine optomechanical imaging systems.

Digitally fabricated instruments allow agile modification and implementation of new imaging techniques without requiring complete system redesign. Design choices that prioritize modularity, interface standardisation, and local adaptability[5] therefore play a central role in accelerating method development while containing complexity. In this article, we define complexity as comprising three aspects: replication complexity (dependence on suppliers, specialized tools, and expertise), assembly and alignment complexity (including documentation quality, part count, and the number of calibration steps), and lifecycle complexity (such as maintenance, optomechanical drift management, and upgrade pathways). In optical imaging systems, where precision optics must integrate with mechanics, electronics, software, and often fluid handling, these architectural decisions are crucial for enabling reliable reproduction and derivation across laboratories.

# Open microscopy as a case study

In this article, we focus on the engineering principles of optical imaging instruments that drive the cutting edge of method development and accelerate global adoption, using open microscopy[10] as a particularly transparent case study. Because open microscopy platforms share design files, component lists, and fabrication strategies, they provide a transparent testbed for design decisions and systematic examination of how engineering decisions affect optical performance, replication feasibility, modification pathways, and long-term maintainability. Modern microscopes are complex systems that integrate precision optics with mechanics, electronics, software, and often fluidic subsystems for high-throughput quantitative data generation. Developers of open microscopes pursue a variety of objectives, with designs that differ significantly in complexity and cost. Goals include facilitating specialised applications, fostering innovation through adaptable designs, democratizing access to instruments and diagnostics, ensuring affordability, and supporting education and public engagement. The proliferation of open-source microscopy projects in recent years has driven innovation in imaging modalities[10], providing a valuable lens into how open engineering practices can advance high-performance optical imaging instrumentation through distributed development.

Our survey of publicly shared microscopy projects reveals heterogeneity in the field. Because the use of keywords and the in-text identification of the methods under review are highly inconsistent, and in many cases entirely absent, conducting a systematic review of this topic is impractical. Consequently, as a sample, we reviewed a community curated list of 80 open microscopy projects on GitHub[11]. Approximately 20% of these designs (16 projects) are microscopes based on the OpenFlexure[12,13] or UC2[14] ecosystems which rely on low-cost 3D printing[8] and are frequently used for new designs. Among the remaining 64 initiatives, about half (33 projects) do not appear to use any digital fabrication, a third (21 projects) incorporate digital fabrication for some components, and a sixth (10 projects) fabricate as many optomechanical parts as feasible themselves. This list is not exhaustive, and there are further non-open microscopy platforms aiming for guided replication without sharing their complete designs, to which our fabrication considerations also apply. We observe that digitally fabricated components are already widely used in open microscopy designs, but their integration is not yet systematic.

Despite widespread use of 3D printed components in open microscopy, there remains limited clarity regarding when and how desktop printing and related methods are appropriate in precision optomechanical contexts. Contrasting positions in the recent literature illustrate this uncertainty. Some developers emphasize the flexibility and performance achievable through geometry optimisation and composite design[15,16], whereas others prioritise conventional scientific suppliers and standardised components because of their convenience and established performance[17]. These differences reflect not a fundamental disagreement about optical requirements, but a lack of shared engineering criteria for assessing when digitally fabricated optomechanical parts satisfy the adoption, innovation, and quality needs required for research-grade imaging. Digital fabrication can be performed directly by research labs or local service providers and contrasts with an increased reliance on specialised scientific providers. Budget, supplier availability, and delivery times vary substantially between regions, creating replication barriers that may not be evident to developers operating within well-resourced supply chains. Designs that depend on specific commercial suppliers or tightly defined component ecosystems can therefore introduce hidden replication complexity and logistical bottlenecks, limiting how readily optical systems and associated protocols can be reproduced or adapted across laboratories[5].

Across projects, we found that digital fabrication most commonly involved FDM 3D printing of holders for optical and structural components, typically without advanced functionality (Figure 1, parts highlighted in blue). These holders were often multipart assemblies, sometimes motorised (Figure 1a, c) or incorporating electrical components (Figure 1, all panels). Fully 3D printed microscopes are generally constructed from modular units, occasionally adhering to standardised formats, such as cube-based systems (Figure 1d). In-house CNC-machined aluminium components, usually anodised by external services, were used for custom connectors or structural reinforcement (orange parts, Figure 1a,b). Some groups also used specialty materials, either printed or machined, to achieve specific mechanical or optical properties (Figure 1a, red; 1e, orange), whereas others designed parts to interface directly with PCBs for compact integration (Figure 1e, right). Laser cutting, although less common, enabled the efficient fabrication of planar holders, flow chambers, and larger enclosures from acrylic or wood (Figure 1f). The variety of strategies underscores the adaptability of digital fabrication tools, but also points to uncertainty about best practices. In the following sections, we aim to clarify these choices and help reduce the underuse or misapplication of digital fabrication in optical imaging instrumentation.

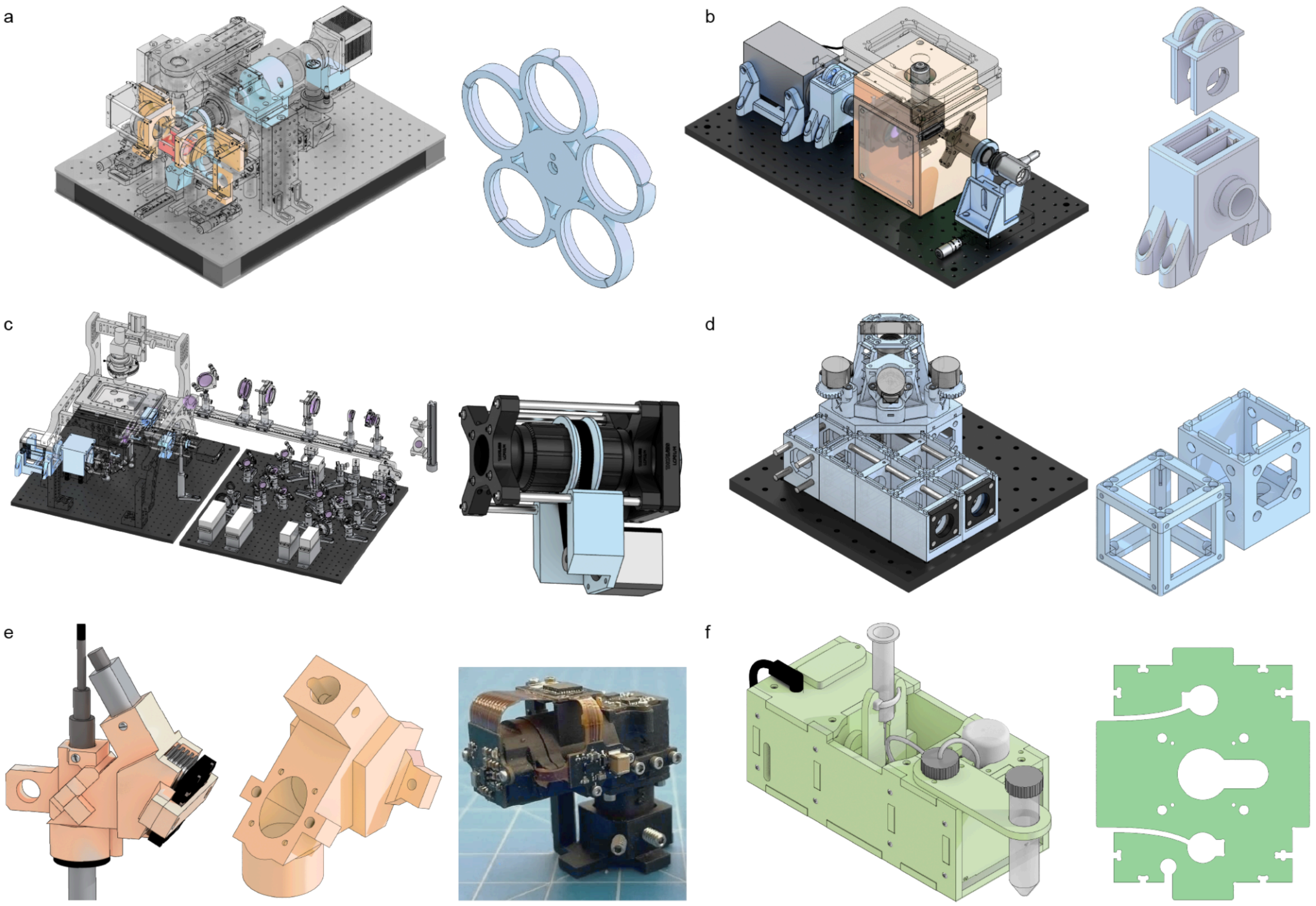

**Figure 1 |** Examples of digital fabrication in research-grade open microscopy, with a systems-view highlighting parts that were FDM 3D printed (blue), custom CNC machined (orange), powder printed (red), and laser-cut (green), with a close-up of an example component. (a) Benchtop mesoSPIM light sheet microscope with filter wheel example[18]. (b) miCube superresolution microscope with astigmatism block[15]. (c) Lifehack microscope for single molecule imaging with motor mount[19]. (d) M4All microscope[20] combining computational alignment, cubes, and the OpenFlexure stage[12]; right: close-up of one of its cubes (back-right), which are similar to the UC2 modular cubes[14], also shown (front-left). (e) Mouse head mounted miniscopes: two-photon Mini2P[21] with machined biocompatible polymer body (left and middle), and Miniscope-LFOV[22] with flex PCBs (right). (f) Portable microscope with fluidics control and a laser-cut body[23].

Dissemination practices also differ substantially. While some “open microscopy”[10] projects provide openly licensed[24,25] design files consistent with Open Source Hardware Association definition[26] or the DIN specification of Open Hardware[27,28], others use restrictive licences, omit formal licensing, or restrict design access, which conflicts with established open hardware standards[27]. These constraints on file access and the sharing of modifications, as well as the lack of permissions to build parts through local fabrication or commercial outsourcing, increase reproduction complexity and cost, and limit the evolution of digitally fabricated imaging platforms.

## Digital fabrication and optical performance

Our survey and experience suggest that an underused opportunity for digital fabrication currently lies in modularly integrating optical instruments rather than replacing precision optics. Printed parts in reviewed microscopes were mostly holders and structural elements; designed deliberately, they can combine mounting, alignment, motion, and modular interfaces. This enables integration of commercial objective lenses and optics, industrial imaging components, maker electronics, and open control software. Together, these resources allow optical systems to be built, modified, automated, and shared easily without redesigning subsystems. However, digital fabrication affects optical performance through three routes: directly through still-emerging printable optical elements; indirectly through selecting and integrating with application-matched optical components and software; and through printed optomechanics' accuracy and temporal stability in maintaining optical geometry. Therefore, care must be taken to design the system for the imaging specifications required.

Direct fabrication of functional optical elements remains the least mature of these routes. Even though 3D printed custom lenses and other functional optical elements[29–33] have recently achieved useful imaging performance, including subcellular detail, their fabrication and validation remain more specialised than the printing of optomechanical parts. One study also reported declining contrast and resolution as additional printed lenses were introduced with current materials[29].

Because printed optics are not routine, performance in digitally fabricated instruments is largely set by the selection of commercial components. Their cost and performance tradeoffs are application-dependent: numerical aperture, wavelength, and aberration correction constrain optical resolution; pixel pitch relative to magnification determines sampling; sensor dimensions and magnification set the nominal field of view, of which only the area maintaining acceptable image quality is usable; detector noise, quantum efficiency, and dark current affect photon budget and exposure time. For example, benchtop mesoSPIM identified industrial long-working-distance objective lenses as the best performers among 34 candidates after quantitatively comparing contrast and field flatness across the wide field of view[18]. Similarly, an industrial CMOS camera showed poorer dSTORM fluorophore localisation at very low photon counts compared with an sCMOS camera, yet both achieved equivalent reconstructed superresolution[34]. Machine-vision vs scientific cameras should therefore be selected by sensor performance: small pixels may oversample without improving optical resolution, whereas active sensor dimensions limit the captured field, and detector characteristics determine weak-signal and long-exposure performance. When application requirements are met, the lower cost and compactness of many industrial cameras can make additional detectors practical for multimodal acquisition, system characterisation, or image-based calibration and alignment[20].

Open software interfaces then determine how effectively components can be coordinated and corrected. Modular hardware aligns naturally with software-defined experimental logic, enabling feedback loops. Closed-loop control of stages, illumination, fluidics, and other actuators can improve signal acquisition, optical alignment, and enable smart applications. Separating hardware drivers, experimental logic, and adaptive analysis facilitates automation, parallelisation, and long-term operation, particularly in time-lapse, incubated, or perturbation-based workflows[35]. Software-coordinated acquisition and reconstruction can additionally improve system-level imaging performance through computational imaging approaches[36].

Once suitable components have been selected, printed optomechanics determine how accurately and stably their required relative positions are realised. Dimensional variation, warping, accumulated interface tolerances and clamping loads can alter lens decentring, tilt and spacing, or deform thin optics. Depending on the optical design, this may produce field-dependent aberrations, degradation of the point-spread function and contrast, or optical-axis displacement. These effects cannot be inferred from printer resolution alone because they depend on the affected element and the sensitivity of the complete optical system. Digital fabrication also allows interfaces to be readily redesigned: in one printed laser system, refinement of the lens-seat geometry reduced measured peak stress in the lens from 1.04 to 0.34 MPa, equal to the unmounted lens and below the 0.48 MPa measured in a commercial retaining-ring mount, although the resulting change in wavefront quality was not measured[37].

Full optical tolerance analysis is not required for every printed part. A useful first screen is to identify whether the part defines a critical optical relationship and compare its expected error with an application-relevant scale: axial motion with the permitted focus error or axial point-spread function; lateral motion with the field, tracking, tiling or registration requirement; and angular error with the displacement it produces along the optical path. Because the depth of focus scales inversely with the square of the numerical aperture (NA) used, a 10 µm displacement may be unacceptable at high NA focusing but inconsequential at low NA. Detailed sensitivity modelling, ray tracing, or point-spread function (PSF), modulation transfer function (MTF) or wavefront measurements are useful where this initial comparison indicates a small performance margin or where phase and beam quality are themselves critical.

Stability over time is likewise a property of the assembled system and its operating conditions. Long-term measurements of the monolithic OpenFlexure stage showed less than 20 µm lateral displacement drift over the course of a week and less than 10 µm relative objective–sample axial drift over five days under ambient conditions[13]. An ESPressoscope prototype showed lateral variation of approximately ±4 µm at 20 °C and 40% relative humidity, whereas an incubator experiment exhibited an initial displacement of approximately 200 µm, predominantly laterally, during the first 30 min and a net lateral displacement of approximately 100 µm after three days[38]. In other incubated microscopes, using ABS or ASA rather than PLA for highly loaded stage components, and allowing thermal equilibration enabled seven-day acquisitions without automatic or manual refocusing[14,39]. These measurements should not be interpreted as universal material-specific thresholds: axial and lateral motion must be judged against the focus, field and registration requirements of the intended acquisition.

System performance must be evaluated against the intended application. Individual parts differ in their relevance to the optical system: a part with loose tolerances can be acceptable if it does not define a critical optical relationship. Where passive accuracy or stability is insufficient, three strategies can be combined. Alignment adjustment can compensate for dimensional deviations of printed parts. Selective use of high-precision components at critical interfaces concentrates precision where it is needed without raising the complexity of the whole system. Image-based methods can correct alignment and defocus dynamically[20,39,40]. Table 1 summarises the principal failure modes and the corresponding validation tests, Supplementary Table 1 provides application-defined acceptance checks, screening relations and test methods for each optical outcome, and Supplementary Table 2 gives material, fabrication and design options. The relevant comparison is therefore between complete architectures and the optical functions they must maintain, rather than between printed polymer and machined metal parts in isolation.

**Table 1** | Issues and validation routes in digital fabrication of optical systems. System performance must be evaluated towards a specific application; printer specifications alone do not directly establish optical adequacy. Expanded screening relations and test methods are provided in Supplementary Table 1. PSF, point-spread function; MTF, modulation transfer function.

| Optical outcome | Fabrication-related risks | Required test |
|---|---|---|
| Imaging quality | Low dimensional precision, thermal expansion, mechanical creep and outgassing. | Measure PSF, MTF, contrast as a function of field of view. |
| Excitation path or beam alignment quality | Thermal expansion, mechanical creep, low dimensional precision, beam clipping, low stiffness. | Measure beam profile, power efficiency in different points along the path and potential wavefront errors. |
| Stability | Thermal expansion, mechanical creep, point-contact wear, low stiffness, anisotropic mechanical properties, hygroscopic swelling. | Track fiducials in 3D over time, temperature, and humidity. |

# Application highlights

Ultimately, the value of digital fabrication in optical imaging systems lies in its ability to deliver robust, reproducible, and scientifically relevant data. Digitally fabricated and self-built systems enable close co-design of hardware, control, and data workflows through modular architectures and accessible interfaces. As a result, performance and experimental flexibility arise not from hardware alone, but also from its tight coupling with programmable control, automation, and data processing. Several systems with a high proportion of 3D-printed components have successfully supported advanced quantitative imaging applications. For example:

- The OpenFlexure platform imaged *Plasmodium* parasites in red blood cells[41] using brightfield detection (Figure 2a), demonstrating reproducible sample preparation and analysis pipelines suitable for diagnostic workflows. Furthermore, 3D printed devices have been shared to reproducibly generate high-quality blood smears[42,43].
- The M4All Dual Channel Fluorescence microscope was developed to enable studies of macrophage host defense mechanisms in the context of bacterial internalisation[44,45] (Figure 2b).
- Combining 3D printed hardware with computational microscopy techniques can enhance the functionality of these systems. One example is the use of the transport of intensity equation to generate computational quantitative phase maps of samples from brightfield images. This technique has been employed to image details of algal cells[46] (Figure 2d,e).
- An open-source microscope based on the UC2 toolbox enabled dSTORM super-resolution imaging of microtubules using a hybrid of injection-moulded and 3D-printed components[47] (Figure 2f,g).
- The Picroscope system supported stable longitudinal live imaging of *Xenopus tropicalis* embryo development for over 28 h inside an incubator[48] (Figure 2h). Since longitudinal studies monopolise instruments during acquisition, this data is often difficult to obtain in shared facilities and particularly benefits from low-cost parallelisation and automated acquisition pipelines.

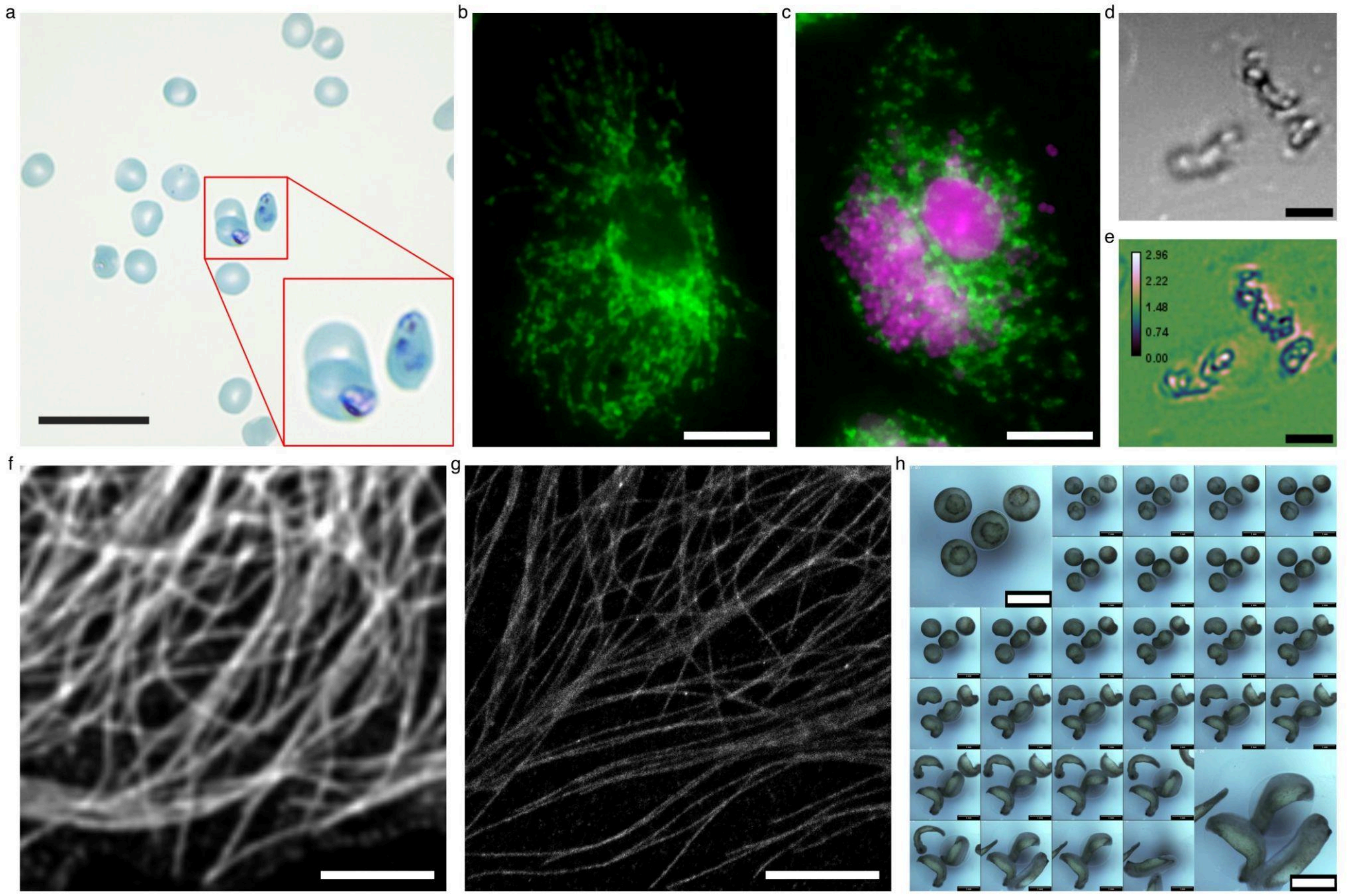

**Figure 2** | Examples of research-grade bioimaging applications using different 3D printed microscopes. (a) Brightfield detection of malaria parasites within red blood cells using the OpenFlexure Delta Stage[41]. Scale bar: 20 µm. (b,c) Widefield fluorescence image of human monocyte-derived macrophages acquired using an M4All Dual Channel Fluorescence microscope[45]. Scale bar: 10 µm. (b) healthy control cell with mitochondria (green) labelled with anti-TOMM20 and Alexa Fluor 488. (c) Mitochondria (green) after bacterial internalisation of *Staphylococcus aureus* (magenta), labelled with DRAQ5, which also stains the nucleus of the cells. (d,e) Images of Thalassiosira pseudonana using the OpenFlexure Microscope[46]. Scale bars are 10 µm. (d) Brightfield image and (e) corresponding computational quantitative phase map. (f,g) Microtubules imaged using an open-source microscope developed with the UC2 toolbox[47]. Scale bars: 3 µm. (f) Widefield fluorescence mode. (g) Super-resolution dSTORM image. (h) Longitudinal live imaging of *Xenopus tropicalis* embryo development over 28 h using the 3D-printed Picroscope[48]. Scale bars: 1 mm (white).

Across these systems, digital fabrication enables modular hardware architectures that integrate naturally with programmable control, automation, and data analysis, allowing smart and adaptive experimental workflows without redesigning entire instruments. These use cases demonstrate that digitally fabricated systems, when thoughtfully designed and appropriately benchmarked, can support advanced imaging experiments by combining hardware with software-driven control and analysis. Among others, they enable high-quality phenotypic analysis, studies of infectious disease mechanisms, and dynamic developmental biology. Although the examples here focused on bioimaging, the same design strategies can be applied to a broader class of analytical instruments, such as those for spectroscopic analysis, temperature dependent analysis, fluidic control, and motion actuation. Digital fabrication can, when applied strategically, increase the impact of instrumentation designs by making them more adaptable, parallelisable, and accessible.

# Choosing a platform and fabrication strategy

The application examples above show that demanding imaging workflows can be realised through different combinations of printed, moulded, machined, and purchased components. The relevant decision for a given application is therefore if an existing platform can be adopted or extended and which functions benefit from local fabrication to support dissemination and innovation speed. Extending an existing platform design is generally preferable when it meets the scientific workflow because documentation, validation, and community experience can accumulate, especially in open projects. Routine users contributing to development can result in particularly practical designs. A new platform or subsystem is justified when existing designs cannot meet the required functionality or target users. Digital fabrication is particularly valuable when it enables otherwise difficult functions such as metal-free designs for magnetic sample manipulations, lightweight structures for in-vivo analysis, or rapid prototyping to support iterative assay development, improving accessibility and adaptability, and reducing fabrication time. 3D printing enables complex part geometries that may combine features to reduce the number of parts required; in turn reducing the assembly time and alignment errors that directly affect experimental stability and reproducibility. This assessment should include sourcing, design, assembly, calibration, validation, maintenance, and modification rather than nominal part price alone.

Implementing a published instrument design is not necessarily plug-and-play. Textbox 1 makes the practical recommendations for adoption explicit. A project that does not meet most criteria may remain a useful design reference, but the missing elements represent development work that the adopting laboratory must plan and resource. Where possible, extensions and corrections should be contributed to the originating project as editable, versioned files and documentation.

**Textbox 1** | Six tips to evaluate when adopting instrumentation resources. Among the growing number of available instrumentation projects to build, it can also be challenging to identify suitable designs for reuse or extension. Even though "reinventing the wheel" may receive academic credit in some cases[49], contributing to existing design efforts is generally more impactful and more sustainable.

**Six tips to evaluate when adopting instrumentation resources**

1. **Documentation**: Look for comprehensive, publicly available, build instructions. Where possible, engage by contributing questions or suggestions to project repositories to support their continuous improvement. Projects lacking documentation may still be useful as design references but require more time and expertise for adoption.
2. **Community and activity**: Check for recent updates, responsiveness (e.g., GitHub or GitLab issues), and external validation through replication or peer-reviewed journals. An activity signals ongoing support.
3. **File availability**: Ensure that source design files, particularly original parametric CAD models, are shared alongside printable formats (e.g. STL), as the latter are difficult to adapt or modify
4. **Fabrication**: Assess whether key components can be fabricated locally, accessed via on-demand manufacturing services, or are commercially available at your location. Designs that rely on hard-to-source components can limit global accessibility.
5. **Benchmarking**: Confirm whether the instrument or subsystem has been tested in realistic applications or biological contexts. Benchmarks should be relevant to your use case.
6. **Expertise**: Can you access the necessary skills or collaborations to build, set up, and calibrate the design?

When modification or a new build is required, module fabrication choice should be justified based on its advantages. FDM is particularly useful for rapid iteration, low-volume complex parts, specialised adapters, modular inserts, and distributed replacement parts. Laser cutting is efficient for suitable planar structures and enclosures; resin printing can provide finer features than typical FDM; CNC machining or commercial metal components may be preferable for stiff, load-bearing, wear-prone, or precision-interface components; and injection moulding becomes attractive when a stable design and repeated production justify tooling. Additionally, integrating widely available industrial or maker components, particularly for structural elements, control and automation, can enhance reliability and functionality while offering a better commercial availability globally and lower cost.

Note that currently, 3D printing is particularly well-suited for distributed manufacturing[1,3], where designs must be reproducible across laboratories and experimental contexts with minimal specialised tooling. It excels at enabling modularity, rapid iteration, and sharing of adaptations; this characteristic makes it ideal for distributing replacement parts and providing specialised adapters, even for traditional industrial instruments. However, for repeated production at a larger scale or demanding mechanical loads, alternative fabrication methods may be preferable, see above and Supplementary Table 2.

It is also important to approach this process with the right motivation: fabricating your own parts with digital fabrication adds steps to a build, and opting for them should reflect an interest in customisation and dissemination, not solely cost-saving. The fabrication method of the project you engage in should therefore align with your goals and capabilities. Weighing the extra time required for design and assembly of custom components against experimental agility, increased iteration speed, and reproducibility across laboratories worldwide helps determine whether digital fabrication offers the most effective route for a given project.

The flowchart in Figure 3 aids project selection and component design decisions for instrumentation users and developers. Once FDM has been selected for a component, success depends on designing around its geometry, mechanics, and replication constraints rather than treating it as a direct substitute for a machined part. The following section addresses these design principles.

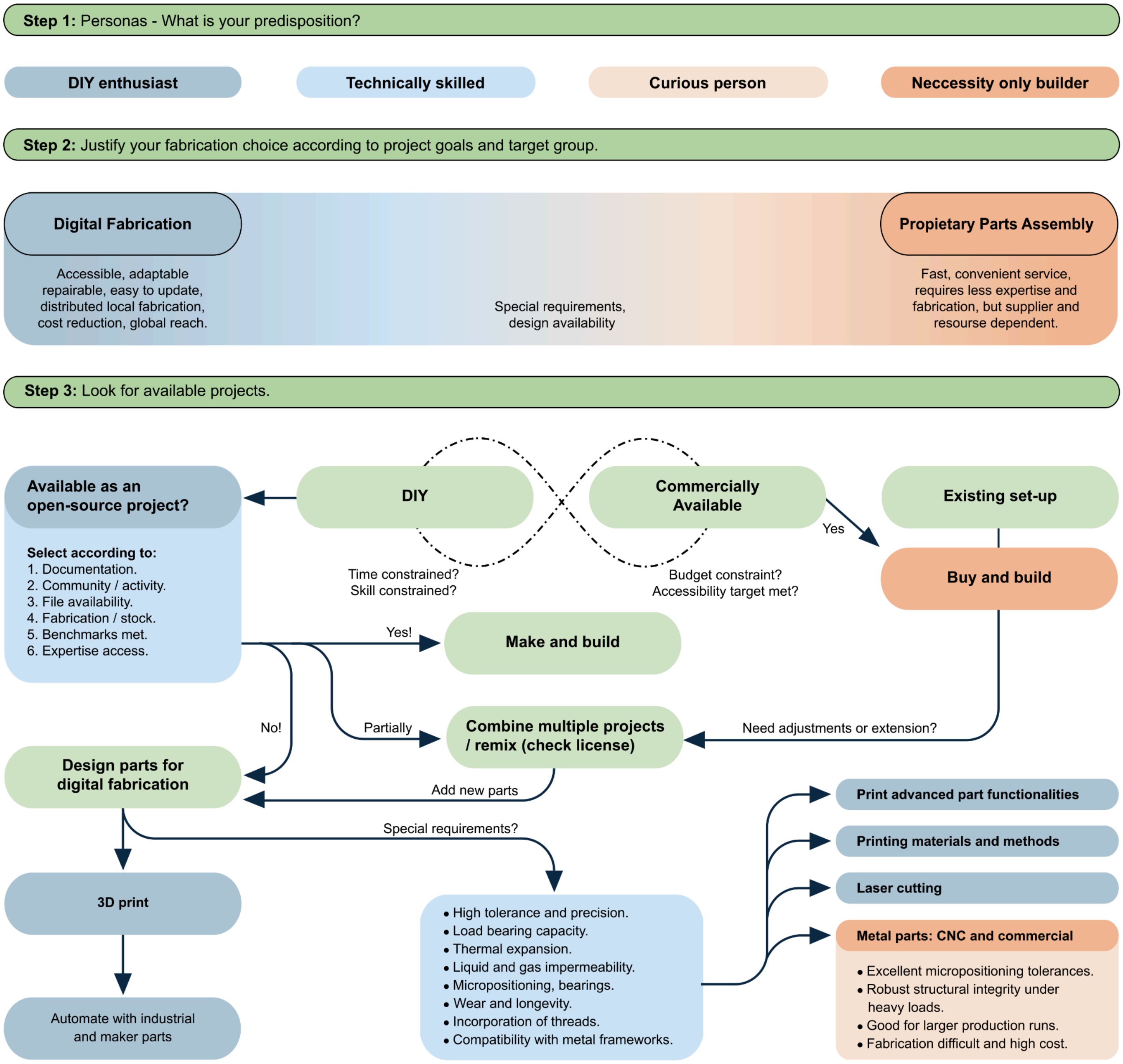

**Figure 3** | Flowchart for imaging platform selection and extension part design, reflecting the different sections of the article. Step 3 includes an abbreviated open-source project selection guide, and the columns of Supplementary Table 2, which details special project requirements with digital fabrication tips and alternatives. When designing parts for digital fabrication, especially 3D printing, refer to the section below for details. In the future, as tools improve and communities grow, this chart will evolve to reflect an increasing DIY ability to design and build more ambitious tools (see section "From fabrication pipelines to scalable infrastructure").

# Design principles for 3D-printed optomechanics

Common FDM materials rarely match precision-machined components in intrinsic dimensional accuracy, stiffness or thermal stability, but digital fabrication provides greater freedom to determine where precision is required and how it is achieved. Once filament 3D printing has been selected as fabrication method, application appropriate optical requirements (Table 1) must be translated into geometry, interfaces, and acceptance tests. Functional accuracy is not determined by a single printer-resolution value; it depends on feature geometry, build orientation, material, slicing, machine condition, and assembly. Effective designs therefore make most fabrication variability functionally unimportant, define a small number of critical relationships, and concentrate adjustment or hybrid precision only where the application requires it. Directly copying a machined metal component design into polymer seldom exploits these possibilities as their fabrication constraints differ[8].

i) **Design around process variability.** Nozzle width and layer height describe the scale of material deposition, not a guaranteed dimensional tolerance. Locating surfaces should be kept free of support scars, and load-bearing features oriented with respect to layer direction. Critical holes, slots, and mating parts should incorporate deliberate clearance or a specified finishing step rather than rely on nominal CAD dimensions. Tapered lead-ins, split clamps, shims, and adjustment slots can accommodate moderate dimensional variation. For sensitive fits, a test piece reproducing the actual interface can be printed with the same material, orientation, and settings. Dimensional accuracy varies across printing technologies, vendors, and printer models[50,51]. Designs intended for reproduction across sites should therefore parameterise function-critical clearances and verify them on each local printing setup. Only features affecting alignment, loading, motion, or serviceability normally require this level of testing.

ii) **Control alignment and tolerance accumulation.** Individual 3D printed optomechanics with a critical optical relationship should reference a short and explicit chain of well-defined reference surfaces “datum” for placement accuracy. Integrating several features into one print can also eliminate uncertain assembly interfaces and combine mounting, alignment, cable routing, and motion. Adjustable optical mounts should preferably use separate features to establish axial position, lateral centring, and retention so that tightening does not shift the component. Defined end stops and self-locating or compliant centring features can improve assembly repeatability, while adjustment should be provided only along optically sensitive degrees of freedom. Monolithic construction is useful when it removes tolerance accumulation, but large prints remain susceptible to warping and can hinder replacement. Modular or hybrid interfaces are preferable where build size, environmental stability, or serviceability dominates. Several projects have proposed a metal-machined frame to mount components, such as openFrame[52], miCube[15] (Figure 1b), and the proprietary ASI RAMM frame in the lifehack microscope (Figure 1c); however, it remains to be seen whether one of them will be more widely adopted. If stable passive location of components is insufficient, the system can incorporate active sensing and actuation.

iii) **Use compliance and flexures deliberately.** Compliant structures[53] can absorb fabrication variation and provide clamping or guided motion without relying on rough printed sliding surfaces, thereby reducing friction, backlash, and part count. Clamps for tight fits, snap fits for movable joints, screw-actuated flexures, and parallelogram stages illustrate this strategy (Figure 4). Lever mechanisms can additionally reduce actuator motion; the OpenFlexure Block Stage demonstrates that an appropriately designed monolithic printed flexure can achieve sub-100-nm positioning resolution[54]. Flexures require smooth transitions and appropriate layer orientation. Printed polymers also creep or lose preload under sustained deformation. Long-term position or clamping force should therefore not normally depend solely on a permanently deflected printed element. A screw, metal spring, magnet[38], elastomer band[13], or actuator can establish the bidirectional force instead.

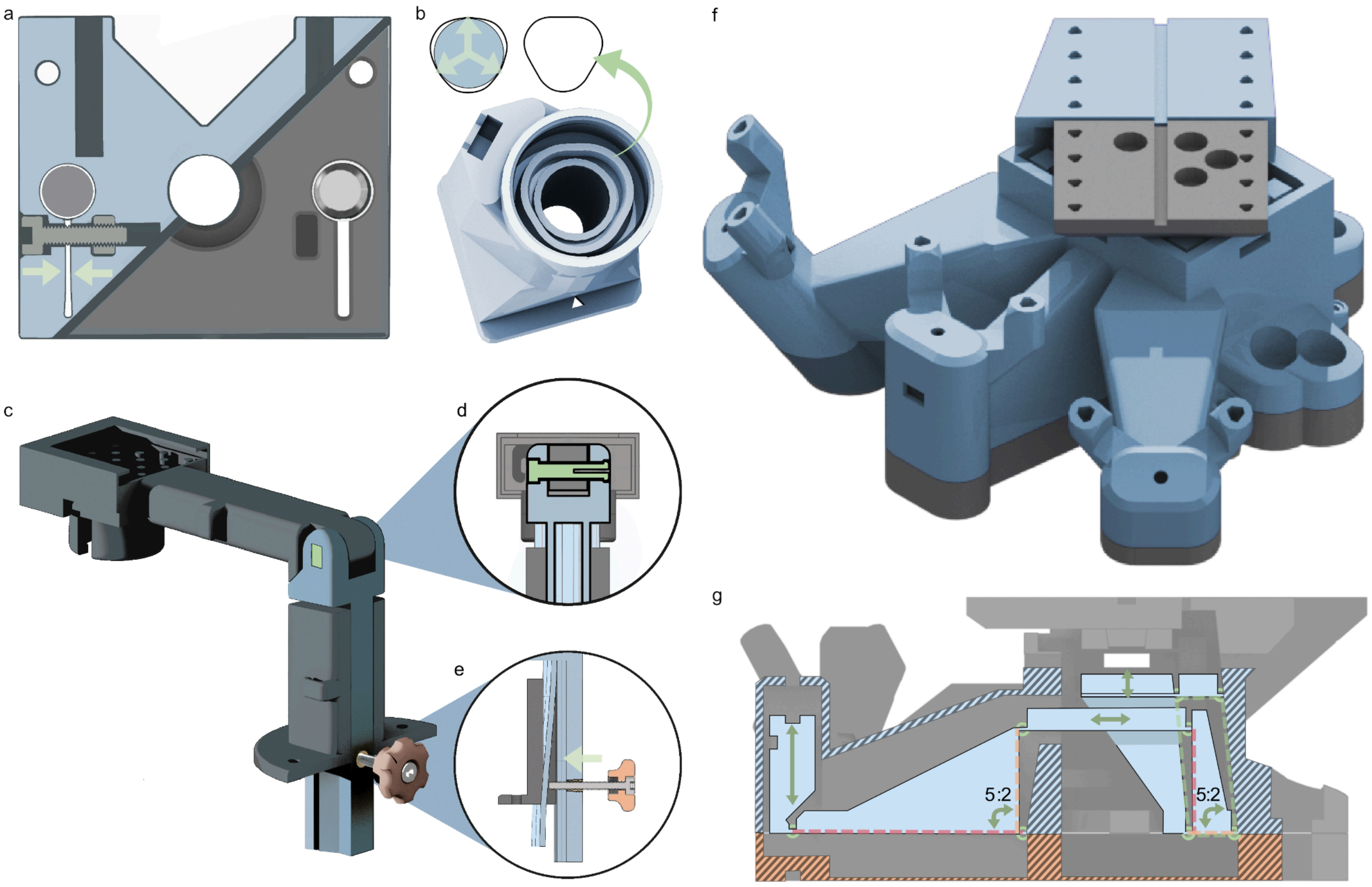

**Figure 4** | Examples of flexure mechanisms enhancing the functionality of 3D printed components. (a) Clamping mechanism securing a 3D-printed part to a metal rod, with an internal screw and nut trap (see blue cutaway) tightening the gap (green arrows)[55,56]. (b) RMS-threaded objective holder with a triangular flexure trap (arrow) securing a tube lens by inward pressure (arrows), keeping it centered and resting flat against an end-stop ring[12]. (c) Folding illumination arm of an inverted microscope, rotating around (d) a joint with a 3D-printed dowel pin (green) that snaps into position[56,57]. (e) The illumination arm is height-adjustable, secured in place by a screw-pressed flexible sliding element (arrow). (f) 3D-printed three-axis fiber alignment stage achieving <100 nm movement precision via a two-stage lever with a total of 5:1 precision increase[54]. (g) Cross section of the same OpenFlexure alignment Block Stage, illustrating the Z-adjustment mechanism: screw-driven motion (left), flexible hinges (green circles), and parallelogram constraints to ensure unidirectional motion (green box) along immovable walls (hatched elements).

iv) **Use durable connections and hybrid precision where needed.** For small, highly loaded, or repeatedly serviced joints, captive nuts or threaded inserts usually provide more durable interfaces than directly printed threads, although performance still depends on the surrounding printed geometry. Coarse printed threads (e.g. RMS or SM1) may be adequate[12,14,56,58] if they are used infrequently, although wear should be monitored. Small metal pins, rods, bearings, plates, or replaceable inserts can similarly provide stable interface datums, wear surfaces, and load paths while the printed body supplies the more complex integrating geometry. Such local hybridisation concentrates precision and durability where they are required without imposing the same manufacturing standard on the entire structure.

v) **Design for the operating environment and verify the assembly.** Material and build orientation should be selected for the sustained load and relevant humidity, chemical, and thermal requirements, including nearby heat sources such as motors or illumination. PolyLactic Acid (PLA) filament is widely used owing to its ease of printing and good dimensional accuracy; however, it is brittle and softens at relatively low temperatures (~60 °C)[59]. PETG, ABS, and ASA offer improved strength and heat resistance for use in incubators[14,39], while flexible (TPU) or reinforced (carbon fibre–filled) filaments enable specialised designs. Solvent compatibility depends on the polymer, printed microstructure and exposure conditions[60], and controlled immersion testing has shown substantial chemical-specific differences in retained tensile and impact strength across eight common FDM filaments[61]. UV and hydrothermal ageing can likewise reduce the mechanical strength of printed PLA[62]. However materials alone do not predict component performance in a given environment; part design is also relevant because stiffness, anisotropy, shrinkage, creep, and fatigue also depend on the printing process and layer orientation. Critical dimensions require testing under operating conditions and after equilibration. Material-specific guidance and alternative fabrication methods are summarised in Supplementary Table 2.

Final acceptance should be based on the assembled instrument rather than isolated printed parts. For reproduction across sites, we recommend a five-step qualification workflow: (1) document the printer, material, build orientation and processing settings; (2) identify function-critical interfaces and dimensions; (3) print representative test objects and adjust parameterised clearances where required; (4) assemble and equilibrate the instrument under its intended load and environment; and (5) validate the relevant optical outcomes using Table 1 and Supplementary Table 1. Validation should include warm-up, load, environmental conditions, and the full experimental duration, and be repeated when a critical material, printer or fabrication setting changes. Reproducible sharing additionally requires editable CAD[63], detailed assembly instructions, and the settings, results and acceptance limits from this qualification. If these design and verification strategies cannot satisfy the application requirements, the critical feature should be produced using a different process or isolated in a hybrid component (see Supplementary Table 2).

# From fabrication pipelines to scalable infrastructure

As digitally fabricated imaging set-ups mature, their impact increasingly depends on how fabrication methods, control software, and workflow automation are integrated into coherent engineering pipelines. Reducing cost and complexity in optical imaging does not arise from fabrication alone, but from aligning physical design, control logic, and replication infrastructure into a reproducible system design. Several converging developments are shaping how digitally fabricated optical instruments evolve from isolated builds into scalable research platforms.

Fabricating the physical foundation: While FDM 3D printing remains the most accessible digital fabrication method, complementary techniques such as SLA printing, laser cutting, and CNC milling are becoming increasingly available. Precision CNC milling could support the local fabrication of mechanically stable parts required for long-term biological experiments, automation, and environmental control. However, CNC workflows remain less accessible due to complex control software, material-specific tooling, and steeper learning curves. As control software and compact CNC systems become more DIY-friendly, this modality may substantially increase the robustness and longevity of self-fabricated parts. Importantly, effective design strategies differ significantly between printing and machining and must be considered early in development.

Software for design and simulation: Component development would benefit from more universal, open, and collaborative design software and better file-sharing practices[63]. Optical system modelling also remains challenging, as most professional software is proprietary and optimised for few-lens systems rather than full experimental setups. Open Python-based ray-tracing tools[64–66] offer promising alternatives for calculating parameters such as magnification, field of view, beam diameter along the optical path, and focal lengths, as demonstrated for M4All[20]. Beyond static modelling, tighter integration of control, modelling and analysis can enable adaptive optical systems; for example, single-pixel microscopy has combined programmable illumination and tunable-lens autofocus with physics-enhanced reconstruction to image through dynamically scattering media[67]. Interoperable links between design, simulation, control, and analysis could thereby shorten iteration from design to deployment and improve data quality.

Interoperability and reuse: As instruments become more modular and software-driven, interoperability becomes a central architectural constraint. Shared metadata[68], standards[27], documentation tools[69], and publication formats[70] enable components and subsystems to be recombined across projects. Hardware backbones such as openFrame[52] and communication interface standards[71–73] illustrate how physical and software interoperability can merge smaller developer groups into larger, more impactful ecosystems. Interoperability reduces replication complexity by ensuring that new modules can be integrated without redesigning entire assemblies.

From prototypes to scalable infrastructure: For digitally fabricated imaging systems to transition from laboratory prototypes to widely deployed research tools, architectural clarity must be accompanied by scalable replication pathways. Permissive licensing, commercial services, and cross-laboratory development form a feedback loop that allows designs to transition from experimental prototypes into widely deployed research infrastructure. Open hardware compatible manufacturers, freelance developers, and resellers[5] lower adoption barriers while preserving modifiability[74]. This pipeline depends on licences[25,75] that explicitly allow commercial use, ensuring that outsourcing, adaptation, and scaling remain compatible with open development. Sustained digital infrastructure such as forums and databases (e.g., microscopy.db[76], microlist[77] and the GOSH forum[78]), and targeted funder support (e.g. napari plug-in grants), remains essential for long-term impact.

Taken together, these developments suggest that the future impact of digitally fabricated optical imaging systems depends not only on individual fabrication methods but on the alignment of design, fabrication, control software, and scalable distributed engineering. Digital fabrication serves as the architectural entry point into this pipeline, enabling complexity reduction in replication, assembly, alignment and lifecycle while preserving research-grade performance and accelerating method evolution.

# Discussion

Digital fabrication reshapes optical imaging not primarily by substituting cheaper parts, but by enabling architectural redesign: integrating multiple functionalities into part geometry, reducing part count, simplifying the integration of custom hardware, different components sets and industrial parts, defining stable interfaces for modular functionality, and treating reproducibility and distributed rebuilding as core engineering criteria. These architectural shifts can reduce cost and complexity in assembly steps, tolerance stacking, alignment burden, maintenance overhead, and procurement friction, and preserve research-grade performance when designed appropriately. This supports faster method evolution and more accessible instruments that can be reproduced, upgraded, and sustained across laboratories.

Open microscopy provides an analytically transparent case study because design decisions, fabrication strategies, and replication practices are visible and community-tested. The engineering lessons generalise beyond open projects: many optical systems can benefit when architectures minimize precision dependencies, embed alignment into kinematics and geometry, and isolate high-stability requirements into few critical components. In many systems, hybrid approaches offer the most robust path to performance and reproducibility, with printed structures offering versatility combined with machined datum surfaces, thermally stable elements, or load-bearing components.

These benefits are not automatic. Digital fabrication can introduce failure modes through anisotropy, creep, thermal drift, and variability across printers, materials, and assembly practices. The central design question is not whether to use 3D printing, but where it improves accessibility and functionality without shifting risk into performance-critical tolerances. Projects that treat replication as an important design constraint are more likely to scale solutions beyond a single laboratory and effectively disseminate imaging methods. This can be achieved through explicit interfaces, accessible alignment procedures, modular subassemblies, and validated part substitutions.

Looking ahead, the impact of digitally fabricated optical imaging systems will depend on coherent development pipelines connecting design, fabrication, simulation, control software, and interoperability standards. As control stacks mature and design workflows become more collaborative, instrumentation iteration may further accelerate and support distributed specialisation without increasing complexity. By treating digital fabrication as an engineering strategy for complexity management and replication, optical imaging platforms can become easier to deploy globally, maintain over time, and adapt as imaging modalities and experimental questions diversify.

# Acknowledgements

We thank Johannes Hohlbein and contributors to the openly maintained HohlbeinLab/OpenMicroscopy resource list, which accepts community contributions and provided the starting point for our survey. We also acknowledge the wider open-hardware and microscopy communities whose shared designs, documentation, and practical experience informed this article.

The author group was intentionally assembled to combine first-hand experience across several established open-microscopy and scientific digital-fabrication initiatives, including platform development, independent adaptation and deployment, computational imaging, microfluidics, laboratory automation, and open-hardware design.

This work was supported by the Chan Zuckerberg Initiative funding to the LIBRE hub project (TW and VP), the Agencia Nacional de Investigación y Desarrollo ANID SENTINET CIN250062 (TW), ANID Fondecyt 11231198 (VP) and ANID Fondecyt 1241621 (TW).

# AI Declaration

The authors used ChatGPT, Gemini and Claude to help structure, draft and proofread prose sections and tables, organising the authors' original content without generating underlying perspectives, scientific claims or interpretations. All authors reviewed and edited the manuscript to its final version and take full responsibility for its content.

# Author contributions

Conceptualisation and writing – original draft: All. Project administration and data curation: TW. Visualisation: TW, MH, VP, GSC and BD.

# Competing interests

R.W.B. initiated and leads the OpenFlexure Project, which is now co-developed and maintained by a distributed community. He was formerly a director and shareholder of OpenFlexure Industries Ltd, a microbusiness that sold products based on the project. B.D. co-developed the UC2 platform and is a co-founder, managing director, and shareholder of openUC2 GmbH, which develops and distributes open-source optical and microscopy systems based on UC2. G.S.C. co-led the academic development of the open-source M4All platform and has no financial interest in it. T.W. and V.P. co-lead LIBRE Hub, an academic, grant-funded open-bioimaging initiative, and have no financial interest in the platforms discussed. M.H. and V.S. contribute academic expertise in scientific digital fabrication and open hardware and have no financial interest in the platforms discussed. These academic and professional relationships are disclosed for transparency. The authors declare no other competing interests.

e0198370.

**Supplementary Table 1** | Design-stage screening and assembled-system validation of optical performance in digitally fabricated and hybrid imaging systems. PSF, point-spread function; MTF, modulation transfer function; FOV, field of view; NA, numerical aperture.

| **Optical outcome** | **Fabrication-related risks** | **Design check and analytical relations** | **System-level validation** |
|---|---|---|---|
| 1. Imaging: On-axis resolution and sampling | Uneven clamping stresses optics[37]; printed seats can cause decentration, tilt or spacing errors that add aberration or shift focus and magnification ($M_{eff}$). Directly printed optics add surface-figure and roughness errors[29,31]. Modular camera swaps can alter pixel pitch and sensor area. | Object-space pixel pitch: $p_{obj} = b \cdot p_{sensor} / \lvert M_{eff} \rvert$ with binning factor b. Sampling the widefield Nyquist cutoff requires $p_{obj} \le \lambda / (4 \cdot NA)$[79]. Use non-paraxial sensitivity modeling to identify dominant error sources and, where near diffraction limited performance is claimed, set wavefront or Strehl limits (common benchmark $S \ge 0.8$)[80]. | Image sub-resolution beads, a pinhole, or a resolution target to measure on-axis 3D PSF width, MTF, or Strehl ratio[68,81], deriving Strehl from the PSF peak relative to an ideal PSF, not from width alone[80]. Vary clamping preload at fixed alignment to isolate stress induced astigmatism, and quantify detector noise under weak-signal conditions[34]. |
| 2. Imaging: Field of view and flatness | Mount warpage, laser-cut kerf taper, or datum errors can tilt the sensor or objective (θ) and decenter optics. Thick printed walls obstruct apertures, causing edge defocus, distortion, or vignetting. | Define the required field and its focus, resolution and contrast limits. Check clear apertures across field radius *r* using a paraxial model[64]. For tilt θ (radians), center-to-edge defocus is $\Delta z_{edge} \approx r\theta$, which must remain below the axial depth of focus ($\propto \lambda / NA^2$). | Image a calibrated grid and uniform fluorescent or transmitted-light reference across the full FOV to map best focus, resolution or contrast, distortion, peripheral brightness, and usable area[18,68]. Compare single-focus images against locally refocused planes to separate field tilt/curvature from off-axis aberrations. |
| 3. Imaging: Contrast and background floor | Low-infill walls can leak light, polymers autofluoresce, rough layer lines scatter grazing incidence stray light, and polymer outgassing or debris can contaminate optical surfaces[8]. | Define the acceptable background level. Cross-check polymer excitation/emission spectra against filter sets at operating wavelengths and powers. Ray-trace[64] stray-light paths to design baffles or matte surface treatments. | Acquire dark frames and illuminated blanks at operating wavelength, power, and exposure. Measure signal-to-background ratio while masking or substituting printed parts to isolate background sources, and report conditions and flat-field corrections[68]. |
| 4. Excitation: Beam delivery and phase fidelity | Tolerance stacking, datum play and warm-up can shift sources, fibers, or relays. Thread stiction hinders fine angular mirror tuning (δ), and warped mounts stress excitation optics. | Provide adjustment along sensitive axes. Free-space beam shift over a path *d* is $\Delta x \approx d\Delta\theta$ ($\Delta\theta$ in radians; mirror tilt δ gives $\Delta\theta \approx 2\delta$, so $\Delta x \approx 2d\delta$). Propagate datum tolerances through the optical model to prevent pupil clipping[64], and thermally isolate heat sources[39]. | Verify pupil centering and clearance. Measure wavefront error, beam profile, illumination flatness, and transmitted power along the path and at the objective back focal plane[81]. Track beam and pupil centroid stability during warm-up and actuation[37]. |
| 5. Multi-channel and modular registration | Printed datums, joint play, camera rotation, and kinematic-mount hysteresis can shift optical axes during module swaps, causing inter-channel translation, rotation, or scale errors. | Ensure exact kinematic constraint, verify contact stresses against polymer yield and creep limits, and specify the maximum uncorrected registration error acceptable prior to software alignment. | Use axial magnification checks in infinite-conjugate paths to aid alignment[20], and image multi-color beads to map 3D inter-channel registration across the FOV[81]. Repeat across module removal/refitting cycles, reporting both raw mechanical error and software-corrected residuals. |
| 6. Stability: Passive thermal and creep drift | Polymer creep, joint settling, hygroscopic swelling, and thermal expansion across the objective–sample path, cables, and stages can cause focal and lateral drift[13,38]; objective-sample loop is usually the most sensitive segment. | Keep the objective–sample mechanical path (L) short and stiff[13]. Estimate thermal drift via $\Delta z \approx \alpha \cdot L \cdot \Delta T$ ($\alpha$, thermal expansion coefficient[82]; $\Delta T$, temperature change), ensuring motion stays within focus tolerance or active autofocus range[39,40]. | Equilibrate under final load[14,38], then track stationary 3D fiducials over the full acquisition duration. Record warm-up separately, log ambient temperature, humidity, and load, and report raw drift alongside autofocus residuals[38–40]. |
| 7. Stability: Active stage positioning | Drive compliance, gear backlash, flexural fatigue, and layer-line stiction can cause hysteresis and settling errors after commanded stage moves or position returns. | Define bidirectional repeatability, accuracy, and settling-time budgets from localization, tracking, or stitching tolerances while preserving adequate FOV margin. Small commanded steps alone do not establish reliable positioning. | Under final load, track fiducials through bidirectional step staircases and repeated excursion-and-return cycles across representative travel ranges[13]. Report commanded vs. measured displacement, return-error distribution, settling time, and inter-run repeatability. |

**Supplementary Table 2** | Detailed overview of engineering requirements for microscope parts and fabrication strategies to meet the needs of demanding applications. Values and comparisons are representative rather than process specifications. Performance depends on the equipment, material formulation, geometry and build orientation, process settings, post-processing, conditioning and measurement method. Function-critical features should therefore be verified using the intended fabrication process and assembled-system requirements.

| **Requirements** | **FDM 3D printing opportunity** | **Alternative** |
|---|---|---|
| Dimensional accuracy and surface finish | FDM printed parts typically have dimensional errors on the order of tenths of a millimetre but vary strongly with geometry, orientation and printer. In one multi-printer study[50], material extrusion showed a standard deviation of 0.36 mm across measured features. Therefore, parameterise and test critical clearances; datum surfaces can be machined or provided by inserts.<br>However, the fast turnaround of printed parts is perfect for multiple rounds of prototypes, and the design itself can be modified to achieve high precision depending on the 3D printer[83].<br>Importantly, the flexibility of the material can be used to improve fit on many occasions (see Figure 4). And there is the possibility of embedding magnets for alignment into the 3D printed parts. | CNC-machined parts of commercial optomechanics provide tighter, supplier-specified tolerances and finished datum surfaces and are preferable where concentricity and flatness is critical.<br>SLA printers can reach good tolerance and precision, but not at the level of metal. Photopolymerisation generally produces finer features and smoother surfaces, but accuracy remains dependent on the printer, resin, supports, washing and post-curing.<br>3D printed parts can be used for replica molding. For example a 3D printed positive mold can be replicated in silicone (negative mold) and then used to cast epoxy resin, forming a replica of the 3D printed part. This ensures a precise replication of the piece in a stiffer material. |
| Stiffness, load stability and creep | Effective stiffness depends on geometry, shell thickness, infill and build orientation. Short load paths, ribs, geometry and inserts can concentrate stiffness and preload where required. Critical structures should be tested under loading.<br>Print FDM at high infill (e.g. 80%) for maximum load-bearing capacity, and honeycomb or gyroid as infill[84,85]. For high loads, organic (topology optimised) geometries distribute stress more evenly; many CAD packages include finite element analysis (FEA) to simulate load.<br>Carbon-fibre-reinforced filaments offer substantially higher stiffness and load-bearing capacity[59]. | CNC and commercial metal mounts provide higher stiffness, much lower room-temperature creep and more predictable preload. Hybrid assemblies can restrict metal components to the critical load paths and datum interfaces.<br>Photopolymer resins can provide high initial stiffness and detail but remain viscoelastic, and their properties depend strongly on post-curing and ageing. |
| Micropositioning and bearings | Flexure-based multi-axis precision positioning, direction change mechanisms and small angle rotations (design principle iii; Figure 4).<br>Low-cost open source stages or manual stages are relatively easy to motorise for automated positioning[12]. | CNC milling may be very difficult to use for this specific task, but metal-based components are commercially available, and pre-alignment is convenient due to the tight dimensional tolerances common to scientific suppliers. |
| Liquid and Gas impermeability | This depends on the printing parameters. In general, more perimeters help reduce gas and liquid permeability[86]. Sometimes the extrusion parameter can be tuned to more than 100%, for example, from 102% to 105% to fill all the voids created by the layer-by-layer process.<br>In some cases, a coat of epoxy is required to make the part as gas and watertight as possible. Some products are available on the market for the specific task of making 3D printed parts watertight, for example Dichtol AM Hydro. | CNC milling of single blocks of material, for example, aluminium or even plastics, for example, Polyoxymethylene (POM or Delrin). Being made from a single block of material, they are gas and water-impermeable.<br>SLA printing is a popular option for water-tight applications.<br>Laser-cut PMMA is both gas and water-impermeable if it is used in combination with suitable gaskets. |

| Requirements | FDM 3D printing opportunity | Alternative |
|---|---|---|
| Thermal and environmental stability | All thermoplastics used in FDM printing exhibit thermal expansion[82] and it is generally higher than metals. Sensitive mechanical paths should be kept short, parts equilibrated under operating conditions[61], and critical dimensions re-measured after representative environmental exposure.<br>It is possible to choose an FDM-printable material with the lowest thermal expansion. CF-reinforced filaments seem to be the best option thus far[59,82].<br>PETG, ABS and ASA are suggested for incubator applications because of their heat and humidity resistance.<br>It is important to remember that the thermal expansion can be affected by the print orientation, which is usually larger in the X and Y directions and lower in the Z direction (anisotropic expansion). | Thermal expansion of SLA resins is smaller than FDM thermoplastics. However, they tend to shrink during printing, and shrinkage, water uptake and photoageing are formulation- and post- cure dependent. High-end SLA printers and resins from established vendors are less affected. Resins withstand higher temperatures and pressures, e.g. useful for injection moulds.<br>Metals provide predictable thermal behaviour, while glass, metal, engineering polymers or commercial components are preferable for solvent contact, elevated temperatures, repeated cycling and optical datums.<br>PMMA as commonly used in laser cutters has a lower thermal expansion than FDM thermoplastics.<br>In the case of automated microscopes, thermal expansion can be mitigated by active drift compensation and automated alignment and calibration[39,40]. |
| Wear and Longevity | Printed parts age[62] but are inexpensive to reprint as replacements. More wear-resistant materials like PETG, ASA, and Carbon Fiber reinforced filaments are easily available. | Machined aluminium or commercial metal components for wear surfaces and load paths; metal pins, bearings or inserts can be embedded in printed bodies (design principle iv). |
| Cost, lead time and iteration | Desktop FDM offers low material and equipment costs and local turnaround commonly measured in hours, making it well suited to rapid iteration and low-volume custom parts. Total cost should nevertheless include design time, failed prints, calibration, finishing, assembly and replacement. | Desktop resin printing offers similar fabrication times but adds resin handling, washing, post-curing and consumable costs. CNC machining and laser cutting usually have higher setup costs and lead times but deliver finished, stable parts. Commercial components avoid fabrication but may have unit costs, shipping delays and limited customisability. Injection moulding becomes attractive mainly at larger production volumes. |
| Incorporation of threads | Printed threads only work for larger diameter threads e.g. RMS objective threads[12,14,56,58], but they are less robust than metal threads.<br>Heat inserts for a large variety of threads are widely available for adding threads to FDM 3D prints. They are a great option because, in comparison to nut traps, they do not move later on, and no nuts fall during the handling of the parts.<br>Nuts, including square nuts, can also be captured in nut traps or embedded during printing. | DIY CNC machining may be very difficult for this specific task, as threads often have to be added manually; however, threads in commercial metal components are very robust.<br>SLA printers can print threads at good resolution, or can be carefully threaded after printing using manual threading or tapping tools; however, these threads are not much more robust than other printed threads. |
| Compatibility with metal systems | 3D printed parts can be compatible and interchangeable with metal systems (e.g. cage, rail and thread standards); for less common systems, adapters may need to be designed rather than downloaded. Embedded magnets can provide repeatable placement. | Commercial adapter plates, or CNC machined aluminium adapters where stiffness or a precision interface is required |

# Author affiliations

1 Institute for Biological and Medical Engineering, Pontificia Universidad Católica de Chile, Chile
* tobias.wenzel@uc.cl
2 CIFAR Azrieli Global Scholars program, CIFAR, Toronto, Canada
3 SUPA, School of Physics & Astronomy, University of Glasgow, UK
4 Department of Physics and SUPA, University of Strathclyde, Glasgow G4 0NG, UK
5 Leibniz Institute of Photonic Technology, Germany
6 Department of BioNanoTechnology, Wageningen University and Research, The Netherlands

# Figure image sources

Figure 1:
e) right panel [[22]], Creative Commons Attribution License 4.0 (CC BY),
all other renderings and schematics are our own.

Figure 2:
a) Licensed under CC BY 4.0 [41]

b,c) Licensed under Strathclyde Thesis Copyright, with permission by the author, co-author in this present manuscript[44,45]

d,e) Licensed under CC BY 4.0[46]

f,g) Licensed under CC BY 4.0[47]

h) Licensed under CC BY 4.0[48]

Figure 3:
No images were included.

Figure 4:
All renderings and schematics are our own.